\documentclass[10pt,a4paper]{article}
\title{\textbf{Optimal $2^K$ Paired Comparison Designs for Third-Order Interactions}}
\author{Eric Nyarko\footnote{corresponding author.~E-mail:~\texttt{eric.nyarko@ovgu.de}}\\
University of Magdeburg, \\
Institute for Mathematical Stochastics,\\
PF 4120, D-39016 Magdeburg, Germany}
\usepackage{amsmath}
\usepackage{booktabs}
\usepackage{dirtytalk}
\usepackage{float,rotating}
\usepackage{diagbox}
\usepackage{slashbox}
\usepackage{appendix}
\usepackage{tikz}
\usepackage{url}
\usepackage{natbib}
\usepackage{array}
\usepackage{longtable,pdflscape,caption}
\usepackage{amsmath,mathtools}
\usepackage{amsthm}
\usepackage{siunitx} 
\usepackage{booktabs}       
\usepackage{dcolumn}        
\usepackage[flushleft]{threeparttable}

\usepackage{dcolumn}
\usepackage{lipsum} 
\date{}
\begin{document}


\maketitle

\begin{abstract}
In psychological research often paired comparisons are used in which either full or partial profiles of the alternatives described by a common set of two-level attributes are presented. For this situation the problem of finding optimal designs is considered in the presence of third-order interactions.
\end{abstract}
\vspace{0.5mm}

\section{Introduction}
Paired comparison experiments have received considerable attention in many fields of applications like psychology, health economics, transportation economics and marketing to study people's preferences for goods or services where behaviors of interest involve either qualitative (so-called discrete choice experiments) or quantitative responses (so-called conjoint analysis). A comprehensive introduction to this general area of paired comparison experiments can be found in \citep{grossmann2015handbook,van1987optimal,louviere2000stated}.\par
Typical with paired comparisons, respondents usually evaluate pairs of competing options (alternatives) in a hypothetical (occasionally real) setting which are generated by an experimental design and are represented by a combination of the levels of several attributes. However, in applications situations may arise in which one may be interested in special relations between the attributes (interactions). For example, \citet{elrod1992empirical} considered a study on student preferences for rental apartments and where up to four attribute interactions were of special interest. The corresponding result is well summarized in Table $2$ of their paper. Another strand of work for the case of direct observation that incorporates three attribute interactions using real data in a randomized clinical trial of high-risk mother-baby dyads can be found in \cite{shiao2007second}. Although not much attention has been given to the four attribute interactions in the literature, the former study serves as a motivation for the present paper where any of four of the attributes interact. \par

In applications the choice task imposes cognitive burden when the alternatives presented are specified by too many attributes which has a detrimental effect on the validity of the estimated parametes. In that situation, a way to simplify the choice task is to specify only a few components (attributes) of the alternatives known as partial profiles \citep[e.g., see][]{grasshoff2003optimal,chrzan2010using,grossmann2018practical}. The number of attributes that are presented in this restricted setting is called the profile strength \citep{grasshoff2003optimal}. \par

The aim of this paper is to introduce an appropriate model for the situation of full and partial profiles and to derive optimal designs in the presence of interactions. We consider the case when the alternatives are specified by a common set of two-level attributes. Work on determining the structure of the optimal designs by this two-level situation has been carried out by \citep{van1987optimal1,van1987optimal,street2001optimal} in the case of full profiles in a main effects and first-order interactions setup, and by \cite{schwabe2003optimal} for partial profiles. Corresponding results when the common number of the attribute levels is larger than two have been obtained by \cite{grasshoff2003optimal} and \cite{nyarko2019optimal2} in a first- and second-order interactions setup, respectively, for both full and partial profiles. The two-level situation for the corresponding second-order interactions setup has been investigated by \cite{nyarko2019optimal}. Here we treat the case of third-order interactions and provide detailed proofs. \par

The remainder of the paper is organized as follows. In Section 2 a general model is introduced for paired comparison experiments. Section 3 provides the third-order interactions model for both full and partial profiles. Optimal designs are characterized in Section 4 and the final Section 5 offers some conclusions. All major proofs are deferred to the Appendix.

\section{General Setting}
In any experimental situation the outcome of the experiment depends on some factors (attributes) $K$ of influence. In that setting the dependence can be described by a functional relationship $\textbf{f}$ which quantifies the effect of the alternative $\textbf{i}= (i_{1},\dots,i_{K})$ for $k=1,\dots,K$ of the attributes of influence. Any observation (utility) $\tilde{Y}_{na}(\textbf{i})$ of a single alternative $\textbf{i}= (i_{1},\dots,i_{K})$ within a pair of alternatives ($a=1,2$) is subject to a random error $\tilde{\varepsilon}_{na}$. In this case the observations can be described by a general linear model
\begin{equation}\label{eq:1}
\begin{split}
\tilde{Y}_{na}(\textbf{i})&= \mu_{n}+\textbf{f}(\textbf{i})^{\top} \boldsymbol{\beta}+\tilde{\varepsilon}_{na},
\end{split}
\end{equation}
where the index $n$ denotes the $n$th presentation in which $\textbf{i}$ is chosen from a set $\mathcal{I}$ of possible realizations for the alternative and the corresponding mean response $\mu_{n}(\textbf{i})=E(\tilde{Y}_{na}(\textbf{i}))$, $\boldsymbol{\beta}= (\beta_{1},\dots,\beta_{p})^{\top}$ is the vector of parameters of interest and $\mu_{n}$ is the block or pair effect. Obviously in order to make statistical inference on the unknown parameters more than one observation is presented  to get rid of the influence of the presentation effect $\mu_{n}$ due to a variety of unobservable influences. 
\par
Typical with paired comparison experiments the utilities for the alternatives are not directly measurable. Only preferences can be observed for comparing pairs of alternatives $(\mathbf{i},\mathbf{j})=((i_{1},\dots,i_{K}), (j_{1},\dots,j_{K}))$. Here we assume that the preference is quantified as the difference between utilities $Y_n(\textbf{i},\textbf{j})=\tilde{Y}_{n1}(\textbf{i})-\tilde{Y}_{n2}(\textbf{j})$. In that case the observations are properly described by the linear model 
\begin{equation} \label{eq:2}
\begin{split}
Y_n(\textbf{i, j})=(\textbf{f}(\textbf{i})-\textbf{f}(\textbf{j}))^{\top}\boldsymbol\beta+\varepsilon_{n}, \\
\end{split}
\end{equation}
with settings $\textbf{x}=(\textbf{i}, \textbf{j})$ which are chosen from the design region $\mathcal{X}=\mathcal{I}\times \mathcal{I}$ of possible pairs of alternatives. Here $\textbf{f}(\textbf{i})-\textbf{f}(\textbf{j})$ is the derived regression function and the random errors $\varepsilon_{n}(\textbf{i}, \textbf{j})=\tilde{\varepsilon}_{n1}(\textbf{i})-\tilde{\varepsilon}_{n2}(\textbf{j})$ associated with the different pairs $(\textbf{i}, \textbf{j})$ are assumed to be uncorrelated with constant variance. \par
The quality of the statistical analysis based on a paired comparison experiment depends on the pairs (alternatives) in the choice sets that are presented. The choice of such pairs $(\textbf{i}_1,\textbf{j}_1),\dots,(\textbf{i}_N,\textbf{j}_N)$ is called a design $\xi_N$ of size $N$. The performance of the design $\xi_N$ is measured by its information matrix
\begin{equation}\label{eq:3}
\textbf{M}(\xi_N)=\sum_{n=1}^{N}(\textbf{f}(\textbf{i}_n)-\textbf{f}(\textbf{j}_n))(\textbf{f}(\textbf{i}_n)-\textbf{f}(\textbf{j}_n))^{\top}.
\end{equation}
As a performance measure in a majority of works about optimal designs for paired comparison experiments, we confine ourselves to the $D$-optimality criterion which aims at maximizing the determinant of the information matrix $\textbf{M}(\xi_N)$.
\par
To enhance efficient comparison of designs with different sample sizes we have to make use of the standardized (per observation) information matrices
\begin{equation}\label{eq:4}
\textbf{M}(\xi)=\frac1N \textbf{M}(\xi_N)
\end{equation} 
which are related to the concept of generalized designs as detailed in \cite{kiefer1959optimum}.  \par
It is worthwhile mentioning that the linear difference model considered here can be realized as a linearization of the binary response model by \cite{bradley1952rank} under the assumption of indifference, $\boldsymbol\beta=\textbf{0}$ \citep[e.g., see][]{grossmann2002advances}. Specifically, under this indifference assumption of equal choice probabilities, the Bradley-Terry type choice experiments in which the probability of choosing $\textbf{i}$ from the pair $(\textbf{i}, \textbf{j})$ given by $\exp[\textbf{f}(\textbf{i})^{\top}\boldsymbol{\beta}]/(\exp[\textbf{f}(\textbf{i})^{\top}\boldsymbol{\beta}]+\exp[\textbf{f}(\textbf{j})^{\top}\boldsymbol{\beta}])$, and the probability of choosing $\textbf{j}$ from the pair $(\textbf{i}, \textbf{j})$ given by $1-\exp[\textbf{f}(\textbf{i})^{\top}\boldsymbol{\beta}]/(\exp[\textbf{f}(\textbf{i})^{\top}\boldsymbol{\beta}]+\exp[\textbf{f}(\textbf{j})^{\top}\boldsymbol{\beta}])$ as in the work of \cite{street2007construction}, amongst others can be derived by considering the linear paired comparison model.
In particular, this assumption simplifies the information matrix of the binary logit model which coincides with the information matrix of the linear paired comparison model. This is the approach taken by \citep{grasshoff2003optimal,grasshoff2004optimal,grossmann2015handbook}.

\section{Third-Order Interactions Model}
Usually, in paired comparison experiments one may be interested in both the main effects and interactions of the attributes. For that setting optimal designs have been derived by \citep{van1987optimal,grasshoff2003optimal} and \cite{nyarko2019optimal} in a first- and second-order interactions setup, respectively. In this paper we derive optimal designs for the third-order interactions model. \par
In the present setting, we consider $K$ attributes each at two levels and assume the preferences for the alternatives in a paired comparison experiment. In what follows, we commence with the situation of full profiles where two options (alternatives) are considered simultaneously. In that case the alternatives are represented by level combinations in which all attributes are involved. The first alternative is denote by $\textbf{i}=(i_1,\dots,i_K)$ and the second alternative by $\textbf{j}=(j_1,\dots,j_K)$, which are both elements of the set $\mathcal{I}=\{-1,1\}^{K}$ where $1$ and $-1$ represent the first and second level of each attribute, respectively. Specifically, the choice set $(\textbf{i},\textbf{j})$ is an ordered pair of alternatives $\textbf{i}$ and $\textbf{j}$ which is chosen from the design region $\mathcal{X}=\mathcal{I}\times\mathcal{I}$. 
Note that for each attribute (component) $k$ the corresponding regression functions $f_k$ is just the identiy, $f_k(i_k)=i_k$ for alternatives $i_k\in\mathcal{I}=\{-1,1\}$ (see e.g.  \cite{nyarko2019optimal}).\par

In the presence of up to third-order interactions we consider the model 
\begin{align}\label{eq:full_direct}
\tilde{Y}_{na}(\mathbf{i})&=\mu_n+\sum_{k=1}^{K}\beta_{k}i_k+\sum_{k<\ell}\beta_{k\ell}i_ki_\ell +\sum_{k<\ell<m}\beta_{k\ell m}i_k i_\ell i_m  \nonumber\\
&\qquad~~+\sum_{k<\ell<m<r}\beta_{k,\ell,m,r}i_k i_\ell i_mi_r+\tilde{\varepsilon}_{na}
\end{align}
for the direct response $\tilde{Y}_{na}(\mathbf{i})$ at the corresponding alternative $\textbf{i}=(i_1,\dots,i_K)$ of full profiles. Here $\beta_k$ denotes the main effect of the $k$th attribute, $\beta_{k\ell}$ is the first-order interaction of the $k$th and $\ell$th attribute, $\beta_{k\ell m}$ is the second-order interaction of the $k$th, $\ell$th and $m$th attribute and $\beta_{k\ell m r}$ is the third-order interaction of the $k$th, $\ell$th, $m$th and $r$th attribute. The vectors $(\beta_k)_{1\leq k\leq K}$ of main effects, $(\beta_{k\ell})_{1\leq k<\ell\leq K}$ of first-order interactions, $(\beta_{k\ell m})_{1\leq k<\ell<m\leq K}$ of second-order interactions and $(\beta_{k\ell mr})_{1\leq k<\ell<m<r\leq K}$ of third-order interactions have dimensions $p_1=K$, $p_2=K(K-1)/2$, $p_3=K(K-1)(K-2)/6$ and $p_4=(1/24)K(K-1)(K-2)(K-3)$, respectively. Hence, the complete parameter vector $\boldsymbol{\beta}=(\beta_1,\dots,\beta_K,(\beta_{k\ell})_{k<\ell}^\top,(\beta_{k\ell m})_{k<\ell<m}^\top,(\beta_{k\ell mr})_{k<\ell<m<r}^\top)^\top$ has dimension $p=p_1+p_2+p_3+p_4$. Here the regression functions are given by
\begin{align}\label{eq:10}
\textbf{f}(\textbf{i})=(i_1,\dots,i_K,(i_ki_\ell)_{k<\ell}^\top,(i_ki_\ell i_m)_{k<\ell<m}^\top,(i_ki_\ell i_m i_r)_{k<\ell<m<r}^\top )^{\top}
\end{align}
of dimension $p$, where in $\textbf{f}(\textbf{i})$, the first $p_1=K$ components $i_{1},\dots,i_{K}$ are associated with the main effects, 
the second set of $p_2$ components $i_{k}i_{\ell}$, $1\leq k<\ell\leq K$, are associated with the first-order interactions, the third set of $p_3$ components $i_{k}i_{\ell}i_m$, $1\leq k<\ell<m\leq K$, are associated with the second-order interactions, and the remaining $p_4$ components $i_{k}i_{\ell}i_{m}i_{r}$, $1\leq k<\ell<m<r\leq K$, are associated with the third-order interactions.
\par
Due to the cognitive burden associated with alternatives involving a large number of attributes and its detrimental effect on the validity of the estimated model parametes, it has become a common practice in the literature to hold the levels of some of the attributes constant in the alternatives that are presented within a single paired comparison. These constant attributes are usually set to zero in the choice task and the remaining attributes with potentially different levels constitute the resulting choice set. The profiles in such a choice set are known as partial profiles, and the number of attributes that are allowed to vary in the partial profiles is called the profile strength, denoted as $S$ \citep[see][]{grasshoff2003optimal,kessels2011bayesian}. Here the remaining $K-S$ attributes are not shown and remain thus unspecified. \par 

Now, for partial profiles a direct observation may be described by model \eqref{eq:full_direct} when summation is taken only over those $S$ attributes contained in the describing subset. Note that a profile strength $S\geq4$ is required to ensure identifiability of the interactions. As already pointed out, we introduce an additional level $0$ for each attribute indicating that the corresponding attribute is not present in the partial profile. In this setting a direct observation can be described by \eqref{eq:full_direct} even when one considres a partial profile $\textbf{i}$ from the set 
\begin{equation}\label{eq:12}
\begin{split}
\mathcal{I}^{(S)}=&\{\textbf{i};\ i_{k}\in\{-1,1\} \textrm{ for $S$ components and} 
\\ 
&\qquad  
i_{k}=0 \textrm{ for $K-S$ components}\},
\end{split}
\end{equation}
of alternatives with profile strength $S$.
In particular, $\mathcal{I}^{(K)}=\mathcal{I}^{(S)}$ in the case of full profiles $S=K$.
For general profile strength $S$ the vector of regression functions $\textbf{f}$ and the interpretation of the parameter vector $\boldsymbol{\beta}$ remain unchanged.
\par

The corresponding paired comparison model is thus given by
\begin{align}\label{eq:11}
Y_{n}(\textbf{i},\textbf{j})&=\sum_{k=1}^{K}(i_k-j_k)\beta_{k}+\sum_{k<\ell}(i_ki_\ell-j_kj_\ell)\beta_{k\ell}+\sum_{k<\ell<m}(i_ki_\ell i_m-j_kj_\ell j_m)\beta_{k\ell m} \nonumber\\
&\qquad+\sum_{k<\ell<m<r}(i_ki_\ell i_m i_r-j_kj_\ell j_m j_r)\beta_{k\ell m r}+\varepsilon_n
\end{align}
as before. However, caution is necessary for the specification of the design region in the case of partial profiles. There it has to be taken into account that the same $S$ attributes are used in both alternatives. To thwart this restriction the design region can be specified as
\begin{equation}\label{eq:12}
\begin{split}
\mathcal{X}^{(S)}=&\{(\textbf{i},\textbf{j});\ i_{k}, j_{k}\in\{-1,1\} \textrm{ for $S$ components and} \\
&\qquad\quad  \ i_{k}= j_{k}=0 \textrm{ for $K-S$ components}\}
\subset\mathcal{I}^{(S)}\times\mathcal{I}^{(S)}
\end{split}
\end{equation}
for the set of partial profiles with profile strength $S$.

\section{Optimal Designs}
In the present setting, we derive optimal designs for the paired comparison model \eqref{eq:11} with corresponding regression functions $\textbf{f}(\textbf{i})$ defined by \eqref{eq:10}. Without loss of generality, we define $d$ as the comparison depth which describes the number of attributes presented in which the two alternatives differ satisfying $1\leq d\leq S\leq K$ \citep[see][]{grasshoff2003optimal}. 
\par

For profile strength $S$ the design region $\mathcal{X}^{(S)}$ can be partitioned into disjoint sets 
\begin{equation}\label{eq:13}
\mathcal{X}^{(S)}_{d}=\{(\textbf{i},\textbf{j})\in\mathcal{X}^{(S)};\ i_{k}\neq j_{k} \textrm{ for exactly $d$ components}\}
\end{equation}
of comparison depth $d$. These sets constitute the orbits with respect to both permutations of the active levels $i_k,j_k=-1,1$ within the attributes as well as permutations among attributes $k=1,\dots,K$, simultaneously in both alternatives. 
\par

Note that the $D$-criterion is invariant with respect to those permutations, which induce a linear reparameterization \citep[see][p.~17]{1996optimum}. 
As a result, it is sufficient to look for optimality within the class of invariant designs which are uniform on the orbits $\mathcal{X}^{(S)}_d$ of fixed comparison depth $d\leq S$. 
\par

Denote by $N_{d}=2^{S}{K \choose S}{S \choose d}$ the number of different (ordered) pairs in $\mathcal{X}^{(S)}_{d}$ which vary in exactly $d$ attributes and by $\bar{\xi}_{d}$ the uniform approximate design which assigns equal weights $\bar{\xi}_{d}(\textbf{i},\textbf{j})=1/N_{d}$ to each pair $(\textbf{i},\textbf{j})$ in $\mathcal{X}^{(S)}_{d}$ and weight zero to all remaining pairs in $\mathcal{X}^{(S)}\setminus\mathcal{X}^{(S)}_{d}$.  
We next obtain the information matrix for these invariant designs.
\newtheorem{lemma}{Lemma}
\begin{lemma}\label{lemma1}
Let $d\in\{0,\dots,S\}$. The uniform design $\bar{\xi}_{d}$ on the set $\mathcal{X}_{d}^{(S)}$ of comparison depth $d$ has block diagonal information matrix
\begin{equation*}
\begin{split}
\mathbf{M}(\bar{\xi}_{d})=\begin{pmatrix} h_{1}(d)\mathbf{Id}_{K}& \mathbf{0}& \mathbf{0}& \mathbf{0}\\
  \mathbf{0}&  h_{2}(d)\mathbf{Id}_{K\choose2}& \mathbf{0}&\mathbf{0}\\
  \mathbf{0} & \mathbf{0}& h_{3}(d)\mathbf{Id}_{K\choose3}& \mathbf{0}\\
 \mathbf{0}& \mathbf{0} & \mathbf{0}&h_{4}(d)\mathbf{Id}_{K\choose4}\end{pmatrix},
 \end{split}
\end{equation*}
where \(h_{1}(d) = \displaystyle \frac{4d}{K}\), \(h_{2}(d) = \displaystyle \frac{8d(S-d)}{K(K-1)}\), \(h_{3}(d) = \displaystyle \frac{4d(3S^{2}-6Sd+4d^{2}-3S+2)}{K(K-1)(K-2)}\) and  \(h_{4}(d) = \displaystyle \frac{16d(S-d)(2d^2-2Sd+S^2-3S+4)}{K(K-1)(K-2)(K-3)}\).  
\end{lemma}
Here, $\textbf{Id}_m$ denotes the identity matrix of order $m$ for every $m$. The three functions $h_1(d),h_2(d)$ and $h_3(d)$ are identical to the corresponding terms for the second-order interaction models considered by \cite{nyarko2019optimal}. \par

Generally, invariant designs $\bar{\xi}$ can be written as a convex combination $\bar{\xi}=\sum^{S}_{d=1}w_{d}\bar{\xi}_{d}$ of uniform designs on the comparison depths $d$ with corresponding weights $w_{d}\geq 0$, $\sum^{S}_{d=1}w_{d}=1$. Consequently, for every invariant design the information matrix can be obtained as the corresponding convex combination of the information matrices for the uniform designs on fixed comparison depths.
\begin{lemma}\label{lemma2}
Every invariant design $\bar{\xi}=\sum_{d=1}^S w_d\bar{\xi}_d$ on the set $\mathcal{X}^{(S)}$ has block diagonal information matrix
\begin{equation*}
\mathbf{M}(\bar{\xi})=\begin{pmatrix} h_{1}(\bar{\xi})\mathbf{Id}_{K}& \mathbf{0}& \mathbf{0}& \mathbf{0}\\
 \mathbf{0} &  h_{2}(\bar{\xi})\mathbf{Id}_{K\choose2}& \mathbf{0}&\mathbf{0}\\
  \mathbf{0} & \mathbf{0}& h_{3}(\bar{\xi})\mathbf{Id}_{K\choose3}& \mathbf{0}\\
 \mathbf{0}&  \mathbf{0}& \mathbf{0}&h_{4}(\bar{\xi})\mathbf{Id}_{K\choose4}\end{pmatrix},
\end{equation*}
where $h_{r}(\bar{\xi})=\sum_{d=1}^Sw_dh_r(d)$, $r=1,2,3,4$.  
\end{lemma}
First we consider optimal designs for the main effects, the first-order interaction, the second-order interaction and the third-order interaction terms separately by maximizing the corresponding entries $h_{1}(d)$, $h_{2}(d)$, $h_{3}(d)$ and $h_{4}(d)$, respectively, in the information matrix. The resulting designs are optimal with respect to any invariant criterion for the corresponding subset of the parameter vector $\boldsymbol{\beta}=(\beta_1,\dots,\beta_K,(\beta_{k\ell})_{k<\ell}^\top,(\beta_{k\ell m})_{k<\ell<m}^\top,(\beta_{k\ell mr})_{k<\ell<m<r}^\top)^\top$. To start with, we mention that the following Result~1, Result~2 and Result~3 paraphrase theorems given in \cite{grasshoff2003optimal} and \cite{nyarko2019optimal} for both first and second-order interaction models and translate them to the present setting of third-order interaction models.
\newtheorem{result}{Result}
\begin{result}\label{theorem1}
The uniform design $\bar{\xi}_{S}$ on the largest possible comparison depth $S$ is optimal for the vector of main effects $(\beta_{1}$ $\dots,$ $\beta_{K})^{\top}$.
\end{result}

This means that for the main effects only those pairs of alternatives should be used which differ in all attributes presented subject to the profile strength $S$.
\begin{result}\label{theorem2}
$\mathrm{(a)}$
For $S$ even the uniform design $\bar{\xi}_{S/2}$ is optimal for the vector of first-order interaction effects $(\beta_{k\ell})_{k<\ell}^{\top}$.
\\
$\mathrm{(b)}$
For $S$ odd the uniform designs $\bar{\xi}_{(S-1)/2}$ and $\bar{\xi}_{(S+1)/2}$ are both optimal for the vector of first-order interaction effects $(\beta_{k\ell})_{k<\ell}^{\top}$.
\end{result}

This means that for the first-order interactions those pairs of alternatives should be used which differ in about half of the attributes presented subject to the profile strength $S$.

\begin{result}\label{theorem3}
$\mathrm{(a)}$
For $S=3$ the uniform designs $\bar{\xi}_{1}$ and $\bar{\xi}_{3}$ are both optimal for the vector of second-order interaction effects $(\beta_{k\ell m})_{k<\ell<m}^{\top}$.
\\
$\mathrm{(b)}$
For $S \geq 4$ the uniform design $\bar{\xi}_{S}$ is optimal for the vector of second-order interaction effects $(\beta_{k\ell m})_{k<\ell<m}^{\top}$.
\end{result}

This means that also for the second-order interactions only those pairs of alternatives should be used which differ in all attributes presented subject to the profile strength $S$.
\par

The optimal designs of Results~\ref{theorem1},~\ref{theorem2} and~\ref{theorem3} are the same as in the first and second-order interactions model \citep[see][]{grasshoff2003optimal,nyarko2019optimal}. However, for the third-order interactions we obtain the following result.
\newtheorem{theorem}{Theorem}
\begin{theorem}\label{theorem4}
There exists a single comparison depth $d^*$ subject to the profile strength $S$ such that the uniform design $\bar{\xi}_{d^*}$ is $D$-optimal for the third-order interaction effects $(\beta_{k\ell mr})_{k<\ell<m<r}^{\top}$.
\end{theorem}

This means that also for the third-order interactions only those pairs of alternatives should be used which differ in a portion of the attributes presented subject to the profile strength $S$. In particular, the corresponding values of $d^{\ast}$ from Theorem \ref{theorem4} that are presented in Table \ref{proofofh4binary} were obtained by first calculating the values of $h_4(d)$ and determining the maximum.

\begin{table}[H]
\centering
\setlength\tabcolsep{0pt}
\caption{Values of the optimal comparison depths $d^*$ of the uniform designs $\bar{\xi}_{d^{\ast}}$ for the third-order interactions with $S\leq K$ binary attributes}\label{proofofh4binary}
\begin{tabular*}{\linewidth}{@{\extracolsep{\fill}}
    *{10}{D{.}{.}{4}}
                }
    \toprule
 S             &4      &5      &6         &7           &8         & 9         &        10&          11      &12       \\
d^* &1   &1   &1      &1    &2  &2  &    2&  3&3\\ \bottomrule
\end{tabular*}
    \end{table}

For results relating to the whole parameter vector $\boldsymbol{\beta}$, we note that a single comparison depth $d$ may be sufficient for non-singularity of the information matrix $\textbf{M}(\bar{\xi}_d)$, i.e.\ for the identifiability of all parameters. This can be easily seen by observing $h_r(1)>0$, $r=1,2,3,4$, for $d=1$. But this is not true for all comparison depths as for example $h_2(S)=h_4(S)=0$. In view of Results \ref{theorem1}, \ref{theorem2}, \ref{theorem3} and Theorem \ref{theorem4} no design exists which simultaneously optimizes the information for the components of the whole parameter vector. Therefore we restrict our attention to the $D$-criterion for the whole parameter vector.
\par

For later use we mention that a design $\xi$ with nonsingular information matrix $\textbf{M}(\xi)$ has a variance function of the form $V((\textbf{i},\textbf{j}),\xi)=(\textbf{f}(\textbf{i})-\textbf{f}(\textbf{j}))^{\top}\textbf{M}(\xi)^{-1}(\textbf{f}(\textbf{i})-\textbf{f}(\textbf{j}))$. This variance function plays an important role for the $D$-criterion. 
According to the equivalence theorem by \cite{kiefer1960equivalence} a design $\xi^{\ast}$ is $D$-optimal if the associated variance function is bounded by the number of parameters $p$, $V((\textbf{i},\textbf{j}),\xi^{\ast})\leq p$ for all $(\textbf{i},\textbf{j})\in\mathcal{X}$.
\par
Now, for invariant designs $\bar{\xi}$ the variance function $V((\textbf{i},\textbf{j}),\bar{\xi})$ is also invariant with respect to permutations and, hence constant on the orbits $\mathcal{X}^{(S)}_{d}$ of fixed comparison depth $d$. Denote by $V(d,\bar{\xi})$ the value of the variance function for the invariant design $\bar{\xi}$ evaluated at comparison depth $d$ where $V(d,\bar{\xi})=V((\textbf{i},\textbf{j}),\bar{\xi})$ on $\mathcal{X}^{(S)}_{d}$. The following result provides a formula for calculating the variance function.
\begin{theorem}\label{thrm4}
For every invariant design $\bar{\xi}$ the variance function $V(d,\bar{\xi})$ is given by 
\begin{equation*}
\begin{split}
V(d,\bar{\xi})&=4d\begin{pmatrix}\frac{1}{h_{1}(\bar{\xi})}+\frac{S-d}{h_{2}(\bar{\xi})}+\frac{3S^{2}-6dS+4d^{2}-3S+2}{6h_{3}(\bar{\xi})}+\frac{(S-d)(2d^2-2Sd+S^2-3S+4)}{6h_{4}(\bar{\xi})}\end{pmatrix}.
 \end{split}
\end{equation*}
\end{theorem}
If the invariant design $\bar{\xi}$ is concentrated on a single comparison depth, then this representation simplifies.
\newtheorem{corollary}{Corollary}
\begin{corollary}\label{cor_thrm4}
For a uniform design $\bar{\xi}_{d^{\prime}}$ on a single comparison depth $d^{\prime}$ the variance function is given by  
\begin{equation*}
\begin{split}
&V(d,\bar{\xi}_{d^{\prime}}) \\
&\qquad=\frac{d}{d^{\prime}}\begin{pmatrix}p_{1}+p_{2}\frac{S-d}{S-d^{\prime}}+p_{3}\frac{3S^{2}-6dS+4d^{2}-3S+2}{3S^{2}-6d^{\prime}S+4d^{\prime2}-3S+2}+p_{4}\frac{(S-d)(2d^2-2Sd+S^2-3S+4)}{(S-d^{\prime})(2d^{\prime2}-2Sd^{\prime}+S^2-3S+4)}\end{pmatrix}.
 \end{split}
\end{equation*}
\end{corollary}
Note that for $d=d^{\prime}$, $V(d,\bar{\xi}_{d})=p_1+p_2+p_3+p_4=p$ which recovers the $D$-optimality of $\bar{\xi}_d$ on $\mathcal{X}^{(S)}_d$ in view of the equivalence theorem by \cite{kiefer1960equivalence}. \par
The following result gives an upper bound on the number of comparison depths required for a $D$-optimal design.
\begin{theorem}\label{theorem5}
In the third-order interactions model the $D$-optimal design $\xi^{\ast}$ is supported on, at most, four different comparison depths  $d^*, d^*_1,d^*+1$ and $d^*_1+1$, say.
\end{theorem}
Further the results on parts of the parameter vector of the $D$-optimal design for the full parameter vector $\boldsymbol{\beta}$ may depend on both the profile strength $S$ and the number $K$ of attributes as can be seen by the following result and the numerical examples presented in Table \ref{tab:1}. In particular, for the case $S=K=4$ of full profiles the $D$-optimal design can be given explicitly. It is worth mentioning that the corresponding situation of $S=K=4$ of full profiles can also be regarded as complete interactions \citep[see][Theorem 4]{grasshoff2003optimal}. Here we show that the corresponding result can be given explicitly.
\begin{theorem}\label{theorem7}
If $S=K=4$ then the design $\xi^{\ast}=\frac{4}{15}\bar{\xi}_1+\frac{2}{5}\bar{\xi}_2+\frac{4}{15}\bar{\xi}_3+\frac{1}{15}\bar{\xi}_4$ which is uniform on all pairs with non-zero comparison depth is $D$-optimal in the third-order interactions model.
\end{theorem}
Note that for $S=K=4$ all four comparison depths are needed for $D$-optimality. \par
For $S\geq 5$, intermediate comparison depths $d$ and $d_1$ with corresponding weights $w_d$ and $w_{d_1}$ the numerical results presented in Table \ref{tab:1} were obtained by direct maximization of $\ln(\det(\mathbf{M}(w_{d}\xi_{d}+(1-w_{d})\xi_{d_1})))$ for the corresponding optimal comparison depth $d^*$ and optimal weights $w^{\ast}_{d^{\ast}}$ where $1-w^{\ast}_{d^{\ast}}=w^{\ast}_{d_1^*}$. In particular, by considering the designs $\xi^*=w_{d^*}^*\xi_{d^*}+(1-w_{d^*}^*)\xi_{d_1^*}$ the numerical results show that two different comparison depths $d^*$ and $d^*_1$ may be needed for $D$-optimality. This is verified by the \cite{kiefer1960equivalence} equivalence theorem in Table \ref{tab:2}. Specifically, for various choices of profile strengths $S=5,\dots,12$ and the optimal comparison depths $d^*$ and $d^*_1$, the corresponding optimal weights $w^*_{d^*}$ satisfy the condition $w^*_{d^*} = d_1^*/(d^*+d_1^*)$ for $d^*=[(S+1)/3]$ and $d^*+d_1^*=S+1$.
\begin{table}[H]
\centering
\setlength\tabcolsep{0pt}
\caption{Optimal comparison depths $d^{\ast}$ and optimal weights $w^{\ast}_{d^{\ast}}$ for the $D$-optimal designs $\xi^*=w_{d^*}^*\xi_{d^*}+(1-w_{d^*}^*)\xi_{d_1^*}$}\label{tab:1}
\begin{tabular*}{\linewidth}{@{\extracolsep{\fill}}
    *{9}{D{.}{.}{4}}
                }
    \toprule
  & \multicolumn{7}{c}{$S$} \\
    \cmidrule(lr){2-9}
     &5  & 6    &    7&8&9 &10&11  &12                                                                                  \\
    \hline
    d^{\ast}              &2      &2      &2         &3            &3          & 3         &        4&           4                \\
w_{d^*}^{*}    &0.667    &0.714   &0.750      &0.667    &0.700   & 0.727  &    0.667&  0.692        \\\hline
 d^*_1              &4      &5      &6         &6            &7         & 8         &        8&           9               \\
w^*_{d_1^*} &0.333   &0.286   &0.250      &0.333    &0.300  &0.273   &    0.333&  0.308\\
\bottomrule
\end{tabular*}
    \end{table}
 The optimality of the so obtained designs has been checked numerically by virtue of the Kiefer-Wolfowitz equivalence theorem. In particular, for full profiles $S=K$ the corresponding values of the normalized variance function $V(d,\xi^{\ast})/p$ are recorded in Table~\ref{tab:2} where maximal values less than or equal to $1$ establish optimality.
 \begin{landscape}
\begin{table}[H]
\centering
\setlength\tabcolsep{0pt}
\caption{Values of the variance function $V(d,\xi^\ast)$ for $\xi^{\ast}$ from Table~\ref{tab:1} in the case of full profiles $S=K$ (boldface \textbf{1} corresponds to the values at the  optimal comparison depths $d^\ast$ and $d_1^*$ included in $\xi^*$).}\label{tab:2}
\begin{tabular*}{\linewidth}{@{\extracolsep{\fill}}
    *{13}{D{.}{.}{4}}
                }
    \toprule
  & \multicolumn{10}{c}{$d$} \\
    \cmidrule(lr){2-13}
   K & 1       &    2     &  3 & 4  &  5  & 6    &    7&8&9 &10&11&12                           \\
    \hline
5 &0.938& \mathbf{1}&0.938& \mathbf{1}&0.938&&&&&&&\\ 
6 & 0.850&\mathbf{1}&0.950& 0.950&\mathbf{1}&0.850&&&&&&\\
7 & 0.792&\mathbf{1}&0.982&0.952&0.982&\mathbf{1}&0.792&&&&&\\  
8 & 0.759&0.998&\mathbf{1}&0.954&0.954&\mathbf{1}&0.998&0.759&&&&\\ 
9 &0.693&0.958&\mathbf{1}&0.966&0.945&0.966&\mathbf{1}&0.958&0.693&&&\\
10&0.644&0.925& \mathbf{1}&0.985&0.958&0.958&0.985&\mathbf{1}&0.925&0.644&&\\
11&0.609&0.901&0.999&\mathbf{1}&0.973&0.960&0.973&\mathbf{1}&0.999&0.901&0.609&\\ 
12&0.566&0.860&0.979&\mathbf{1}&0.982&0.963&0.963&0.982&\mathbf{1}&0.979&0.860&0.566\\  \bottomrule
\end{tabular*}
    \end{table}
\end{landscape}
 
\section{Discussion}
For paired comparisons in a linear model without interactions optimal designs require that the alternatives in the choice sets show distinct levels in each attribute subject to the profile strength \cite{grasshoff2004optimal}. Moreover, in a first-order interactions model pairs have to be used for an optimal design in which approximately one half of the attributes are distinct and one half of the attributes coincide subject to the profile strength \cite{grasshoff2003optimal}. In a second-order interactions model both types of pairs have to be used for an optimal design in which either all attributes have distinct levels or approximately one half of the attributes are distinct and one half of the attributes coincide subject to the profile strength and the total number of attributes available \cite{nyarko2019optimal}. Here it is shown that in a third-order interactions model two types of pairs have to be used in which the numbers of distinct attributes are symmetric with respect to about half of the profile strength to obtain a $D$-optimal design for the whole parameter vector. Optimal designs may be concentrated on one, two, three or four different comparison depths depending on the number of the profile strength. The invariance considerations used here can be extended to larger numbers of levels for each attribute.
\vspace{4mm} \\
\textbf{Acknowledgement}\\
This work was partially supported by Grant - Doctoral Programmes in Germany, $2016/2017~(57214224)$ - of the German Academic Exchange Service (DAAD).
\bibliographystyle{apa}     
\bibliography{reference4th}   
\begin{appendices}
\section*{Appendix}
\begin{proof} of Lemma~\ref{lemma1}. The quantities $h_1(d),h_2(d)$ and $h_3(d)$ can be obtained as in \citep{grasshoff2003optimal,nyarko2019optimal}. The quantity $h_4(d)$ can be obtained on similar lines. First note that for the levels $i,j=-1,1$ we have $i^{2}=1$ and $ij=-1$, $(i-j)^{2}=4$ for $i\neq j$. \par
For third-order interactions we consider attributes $k$, $\ell$, $m$ and $r$, say, and distinguish between pairs in which all four attributes are distinct, pairs in which three of these attributes $k$, $\ell$ and $m$, say, have distinct levels in the alternatives while the same level is presented in both alternatives for the remaining attribute, pairs in which two of these attributes $k$, $\ell$, say, have distinct levels in the alternatives while the same level is presented in both alternatives for the remaining attribute two attributes and, finally, pairs in which only one of the attributes, say, $k$ has distinct levels in the alternatives while the same level is presented in both alternatives for the three remaining attributes. Then $i_{k}i_{\ell}i_{m}i_r=j_{k}j_{\ell}j_{m}j_r$ in the first and third case, while $i_{k}i_{\ell}i_{m}i_r=-j_{k}j_{\ell}j_{m}j_r$ in the second and last case. Hence,
\vspace{-1ex}
\begin{equation*}\label{eq:4.50}
\begin{split}
(i_{k}i_{\ell}i_{m}i_r-j_{k}j_{\ell}j_{m}j_r)^{2}=0\quad\mathrm{ for }\quad i_{k}\neq j_{k}, \ i_{\ell}\neq j_{\ell}, \ i_{m}\neq j_{m} \quad\mathrm{and}\quad i_{r}\neq j_{r},
\end{split}
\end{equation*}
\begin{equation*}\label{eq:4.51}
\begin{split}
(i_{k}i_{\ell}i_{m}i_r-j_{k}j_{\ell}j_{m}j_r)^{2}=4\quad\mathrm{ for }\quad i_{k}\neq j_{k}, \ i_{\ell}\neq j_{\ell}, \ i_{m}\neq j_{m} \quad\mathrm{and}\quad i_{r}= j_{r},
\end{split}
\end{equation*}
\begin{equation*}\label{eq:4.52}
\begin{split}
(i_{k}i_{\ell}i_{m}i_r-j_{k}j_{\ell}j_{m}j_r)^{2}=0\quad\mathrm{ for }\quad i_{k}\neq j_{k}, \ i_{\ell}\neq j_{\ell}, \ i_{m}= j_{m} \quad\mathrm{and}\quad i_{r}= j_{r},
\end{split}
\end{equation*}
and
\begin{equation*}\label{eq:4.53}
\begin{split}
(i_{k}i_{\ell}i_{m}i_r-j_{k}j_{\ell}j_{m}j_r)^{2}=4\quad\mathrm{ for }\quad i_{k}\neq j_{k}, \ i_{\ell}= j_{\ell}, \ i_{m}= j_{m} \quad\mathrm{and}\quad i_{r}= j_{r},
\end{split}
\end{equation*}
respectively, where the roles of the attributes $k$, $\ell$, $m$ and $r$ may be interchanged. 
\par

For given attributes $k$, $\ell$, $m$ and $r$ the pairs with distinct levels in the four attributes occur $\left(\begin{smallmatrix}K-4 \\ S-4\end{smallmatrix}\right)\left(\begin{smallmatrix}S-4 \\ d-4\end{smallmatrix}\right)2^{S}$ times in $\mathcal{X}^{(S)}_d$, while those which differ in the three attributes occur $\left(\begin{smallmatrix}4 \\ 3\end{smallmatrix}\right)\left(\begin{smallmatrix}K-4 \\ S-4\end{smallmatrix}\right)\left(\begin{smallmatrix}S-4 \\ d-3\end{smallmatrix}\right)2^{S}$ times in $\mathcal{X}^{(S)}_d$, while those which differ in the two attributes occur $\left(\begin{smallmatrix}4 \\ 2\end{smallmatrix}\right)\left(\begin{smallmatrix}K-4 \\ S-4\end{smallmatrix}\right)\left(\begin{smallmatrix}S-4 \\ d-2\end{smallmatrix}\right)2^{S}$ times in $\mathcal{X}^{(S)}_d$ and, finally, those which differ only in one attribute occur $\left(\begin{smallmatrix}4 \\ 1\end{smallmatrix}\right)\left(\begin{smallmatrix}K-4 \\ S-4\end{smallmatrix}\right)\left(\begin{smallmatrix}S-4 \\ d-1\end{smallmatrix}\right)2^{S}$ times. As a consequence, since the number $N_d$ of paired comparisons in $\mathcal{X}^{(S)}_{d}$ equals $N_{d}=\left(\begin{smallmatrix}K \\ S\end{smallmatrix}\right)\left(\begin{smallmatrix}S \\ d\end{smallmatrix}\right)2^{S}$, for the third-order interactions the diagonal elements $h_{4}(d)$ in the information matrix are given by

\begin{align}\label{eqn:54}
h_4(d)&=\frac{1}{N_d}\left(\begin{smallmatrix}K-4 \\ S-4\end{smallmatrix}\right)\Big(\left(\begin{smallmatrix}S-4 \\ d-3\end{smallmatrix}\right)2^{S+4}+\left(\begin{smallmatrix}S-4 \\ d-1\end{smallmatrix}\right)2^{S+4}\Big)    \nonumber\\
&=\frac{16(S-d)d(d-1)(d-2)}{K(K-1)(K-2)(K-3)}+\frac{16(S-d)(S-d-1)(S-d-2)d}{K(K-1)(K-2)(K-3)}    \nonumber\\
&=\frac{16d(S-d)(2d^2-2Sd+S^2-3S+4)}{K(K-1)(K-2)(K-3)}.
\end{align}
Finally, it can be noted that all off-diagonal entries in the information matrix vanish because the terms in the corresponding sums add up to zero due to the \say{orthogonality} condition for single attributes.
\end{proof}

\begin{proof} of Theorem~\ref{thrm4}.
First we note that the inverse of the information matrix of the design $\bar{\xi}$ is given by
\begin{equation*}
\begin{split}
\mathbf{M}(\bar{\xi})^{-1}=\begin{pmatrix} \frac{1}{h_{1}(\bar{\xi})}\textbf{Id}_{K}&\mathbf{0}&\mathbf{0}&\mathbf{0}\\
 \mathbf{0} & \frac{1}{h_{2}(\bar{\xi})}\mathbf{Id}_{K\choose2}& \mathbf{0}&\mathbf{0}\\
  \mathbf{0} & \mathbf{0}&\frac{1}{h_{3}(\bar{\xi})}\mathbf{Id}_{K\choose3}&\mathbf{0}\\
 \mathbf{0}& \mathbf{0}& \mathbf{0}&\frac{1}{h_{4}(\bar{\xi})}\mathbf{Id}_{K\choose4}\end{pmatrix}.
 \end{split}
\end{equation*}
Hence, we obtain for the variance function
\begin{align}\label{eqn:17}
V((\mathbf{i},\mathbf{j}),\bar{\xi})=&(\mathbf{f}(\mathbf{i})-\mathbf{f}(\mathbf{j}))^{\top}\mathbf{M}(\bar{\xi})^{-1}(\mathbf{f}(\mathbf{i})-\mathbf{f}(\mathbf{j}))\nonumber\\
=&\frac{1}{h_1(\bar{\xi})}\sum_{k=1}^{K}(i_k-j_k)^2 \nonumber\\
&\mbox{}+\frac{1}{h_2(\bar{\xi})}\sum_{k<\ell}(i_ki_{\ell}-j_kj_{\ell})^2  \nonumber \\
&\mbox{}+\frac{1}{h_3(\bar{\xi})}\sum_{k<\ell<m}(i_ki_{\ell}i_m-j_kj_{\ell}j_m)^2  \nonumber \\
&\mbox{}+\frac{1}{h_4(\bar{\xi})}\sum_{k<\ell<m<r}(i_ki_{\ell}i_mi_r-j_kj_{\ell}j_mj_r)^2.
\end{align} 
From the proof of Theorem $2$ in \cite{nyarko2019optimal} \citep[see also][]{grasshoff2003optimal}, it can be seen that the first, second and third sum on the right hand side of \eqref{eqn:17} associated with the main effects, the first-order interactions and the second-order interactions equal $4d$, $4d(S-d)$ and $4d(3S^2 - 6dS + 4d^2 - 3S + 2)/6$, respectively.
\par
For the terms associated with the third-order interactions, we have $(i_ki_{\ell}i_mi_r$ $-j_kj_{\ell}j_mj_r)^2=4$, if $(i_ki_{\ell}i_mi_r)$ and $(j_kj_{\ell}j_mj_r)$ differ in three of the associated four attributes $k$, $\ell, m$ and $r$ or in exactly one of these attributes, and $(i_ki_{\ell}i_mi_r-j_kj_{\ell}j_mj_r)^2=0$ otherwise. For a pair $(\mathbf{i},\mathbf{j})\in\mathcal{X}^{(S)}_d$ of comparison depth $d$ there are $(S-d)\left(\begin{smallmatrix} d \\ 3\end{smallmatrix}\right)$ third-order interaction terms for which the four attributes $k,\ell ,m$ and $r$ differ in exactly three of the attributes, and there are $d\left(\begin{smallmatrix} S-d \\ 3\end{smallmatrix}\right)$ third-order interaction terms for which the four attributes $k,\ell ,m$ and $r$  differ in exactly one attribute. As a result, there are
\begin{equation*}
\begin{split}
&(S-d)\left(\begin{smallmatrix} d \\ 3 \end{smallmatrix}\right)+d\left(\begin{smallmatrix} S-d \\ 3\end{smallmatrix}\right) \\
&\qquad=(S-d)d(d-1)(d-2)/6+d(S-d)(S-d-1)(S-d-2)/6\\
&\qquad=d((S-d)(2d^2-2Sd+S^2-3S+4))/6
\end{split}
\end{equation*} 
non-zero entries (equal to $4$) in the fourth sum on the right hand side of \eqref{eqn:17} and, hence, this sum equals $4d((S-d)(2d^2-2Sd+S^2-3S+4))/6$.
\par
By substituting this results into \eqref{eqn:17} for fixed $K$ and $S$, it can be seen that the value of the variance function  depends on the pair $(\textbf{i},\textbf{j})$ only through its comparison depth $d$ and obtain the formula proposed.  
\end{proof}
\begin{proof} of Corollary~\ref{cor_thrm4}.
In view of Theorem~\ref{thrm4} it is sufficient to note that the representation of the variance function follows immediately by inserting the values of $h_{r}(\bar{\xi}_d)$ from Lemma \ref{lemma1} and $p_r={K \choose r}$, $r=1,2,3,4$.
\end{proof}

\begin{proof} of Theorem~\ref{theorem5}.
Let $\xi^{\ast}$ be an invariant $D$-optimal design with weights $w_d^\ast$ on the comparison depths $d$ for which the variance function $V(d,\xi^{\ast})$ is equal to the number of parameters $p$ for all $d$ such that $w_d^{\ast}>0$. By Theorem~\ref{thrm4} the variance function $V(d,\xi^{\ast})$ is a polynomial of degree $4$ in the comparison depth $d$ with negative leading coefficient. For integer $d$ the variance function $V(d,\xi^{\ast})$ may thus be equal to $p$ for, at most, four different values of $d$. Now, by the \cite{kiefer1960equivalence} equivalence theorem itself $V(d,\xi^{\ast})\leq p$ for all $d=0,1,\dots,S$. Hence, by the shape of the variance function we obtain that $V(d,\xi^{\ast})= p$ may occur only at, at most two adjacent comparison depths $d^*$ and $d^*+1$ or $d_1^*$ and $d_1^*+1$, say, in the interior. 
\end{proof}
\begin{proof}of Theorem~\ref{theorem7}.
For the design $\xi^\ast$ we obtain $h_1(\xi^{\ast})=8/15$, $h_2(\xi^{\ast})=2/15$, $h_3(\xi^{\ast})=1/30$ and $h_4(\xi^{\ast})=1/120$. Inserting this into the variance function of Theorem~\ref{thrm4} yields $V(d,\xi^{\ast})=5d(-1/2d^3+5d^2-35/2d+25)/4$ which results in $V(1,\xi^{\ast})=V(2,\xi^{\ast})=V(3,\xi^{\ast})=V(4,\xi^{\ast})=15$. Hence, the variance function is bounded by the number of parameters $p=15$ which establishes the $D$-optimality of $\xi^{\ast}$ by virtue of the Kiefer-Wolfowitz equivalence theorem. 
\end{proof}

\end{appendices}

%
%

\end{document}